\documentclass[aps,prl,twocolumn,superscriptaddress,bibnotes]{revtex4}

\usepackage{epsfig}
\usepackage{graphicx}
\usepackage{latexsym}
\usepackage{amsmath}
\usepackage{amssymb}
\usepackage{mathrsfs}
\usepackage{graphicx,color,psfrag,epsfig}
\usepackage{amsfonts}
\usepackage{verbatim}
\usepackage{dsfont}

\newcommand\ba{\begin{array}}
\newcommand\ea{\end{array}}
\newcommand\ben{\begin{equation}}
\newcommand\een{\end{equation}}
\newcommand\bea{\begin{eqnarray}}
\newcommand\eea{\end{eqnarray}}
\newcommand{\al}{\alpha}

\newcommand{\ket}[1]{\left|#1\right\rangle}
\newcommand{\bra}[1]{\left\langle#1\right|}

\newcommand{\dslash}{d \hspace{-0.8ex}\rule[1.2ex]{0.9ex}{.1ex}}

\newcommand{\mbh}[1]{{\textcolor{red}{#1}}}

\begin{document}


\title{Covariant Closed String Coherent States}



\author{Mark Hindmarsh}
\affiliation{Department of Physics and Astronomy, University of Sussex, \\Brighton, East Sussex BN1 9QH, UK}

\author{Dimitri Skliros}
\affiliation{Department of Physics and Astronomy, University of Sussex, \\Brighton, East Sussex BN1 9QH, UK}



\date{\today}

\begin{abstract}
We give the first construction of covariant coherent closed string states, which may be identified with fundamental cosmic strings. We outline the requirements for a string state to describe a cosmic string, and using DDF operators provide an explicit and simple map that relates three different descriptions: classical strings,  lightcone gauge quantum states and covariant vertex operators. The naive construction leads to covariant vertex operators whose existence requires a lightlike compactification of spacetime. When the lightlike compactified states in the underlying Hilbert space are projected out the resulting coherent states have a classical interpretation and are in one-to-one correspondence with arbitrary classical closed string loops.
\end{abstract}


\maketitle

The construction of covariant closed string coherent states with an arbitrary distribution of harmonics has been sought after for many years. In \cite{Calucci87} it was realized that the naive covariant construction based on an analogy with the harmonic oscillator does not lead to physical open string coherent states. The analogous closed string construction which is of greater interest phenomenologically is even more constrained due to the additional complications of level matching, and this becomes non-trivial even in lightcone gauge \cite{Blanco-PilladoIglesiasSiegel07}. 

With the recent realization that cosmic superstrings may lead to observational signatures for string theory the necessity of understanding macroscopic string states with a classical interpretation has become of paramount importance. 
Cosmic superstrings are expected to be produced in the early universe at the end of ${\rm D3}$-$\overline{\rm  D3}$ brane inflation in e.g. models with warped throats (${\rm \mathbb{K}L\mathbb{M}T}$) or large compact dimensions  (see e.g. \cite{CopelandMyersPolchinski04,CopelandKibble09b} and references therein). Almost all predictions to date concerning cosmic superstrings are either classical and neglect effects of gravitational backreaction (which can be important even for order of magnitude estimates \cite{QuashnockSpergel90}), involve cosmic strings in their vacuum state (with no harmonics excited) \cite{JacksonJonesPolchinski05}, or involve massive momentum eigenstates (with only first harmonics excited) \cite{MitchellSunborgTurok90,ChialvaIengoRusso03,Iengo06,GutperleKrym06,Chialva09} which are not expected to reproduce the classical evolution \cite{Iengo06}.  These computations need to be extended to more realistic cosmic superstrings  and in what follows we discuss the first construction of a closed string covariant coherent state  \footnote{We call a vertex operator, $V$, coherent provided it is: (1) specified by a set of continuous labels $(\lambda,\bar{\lambda})=\{\lambda_n^i,\bar{\lambda}_n^i\}$; (2) admits a resolution of unity \cite{KlauderSkagerstam85}, $\mathds{1} = \int d\lambda d\bar{\lambda} |V(\lambda,\bar{\lambda})\rangle\langle V(\lambda,\bar{\lambda})|$, with $d\lambda d\bar{\lambda}=\prod_{n,i}\frac{d^2\lambda_n^i}{2\pi n}\frac{d^2\bar{\lambda}_n^i}{2\pi n}$ and $d^2x=i dx \wedge dx^*$; (3) transforms correctly under all string theory symmetries.} with arbitrarily excited harmonics, a large fundamental cosmic string loop.  Further details and the corresponding open string construction will be presented in a companion paper \cite{SklirosHindmarsh09}.

Classically, a cosmic string with position $X^\mu$ depending on worldsheet coordinates $z,\bar{z}$ (see  \footnote{Conventions: $z=e^{-i(\sigma+i\tau)}$ and $\bar{z}=e^{i(\sigma-i\tau)}$ from which $\partial_{\tau}=z\partial+\bar{z}\bar{\partial}$ and $i\partial_{\sigma}=z\partial-\bar{z}\bar{\partial}$. Note that $\tau=(\tau)_{\rm Euclidean}=i(\tau)_{\rm Minkowski}$. We take $\alpha'=2$.}) evolves according to the equations of motion and constraints \cite{GSW1}, $ \partial\bar{\partial} X^{\mu}=0$, $(\partial X)^2=(\bar{\partial}X)^2=0$. Explicit solutions are easily obtained in lightcone gauge where one takes $X^+=2p^+\tau$, and for the transverse directions one finds,
\begin{equation}\label{eq:Xcl}
X^i_{\rm cl}(z,\bar{z})-x^i=-ik^i\ln |z|^2+i\sum_{n\neq0}\frac{1}{n}\,\big(\xi_n^i\,z^{-n}+\bar{\xi}_n^i\,\bar{z}^{-n}\big)
\end{equation}

In string theory cosmic strings are described by vertex operators. These are  composed of the fields present in the theory, $X(z,\bar{z})$ and $g_{\alpha\beta}(z,\bar{z})$. Due to conformal invariance the explicit dependence on $g_{\alpha\beta}$  drops out \cite{Weinberg85,Polchinski_v1}, states in the underlying Hilbert space transform like one-particle states under Poincar\'e transformations \cite{Weinberg_v1}, and therefore normal ordered closed string vertices are of the form:
$$V(z,\bar{z})=\sum_{\alpha} \mathcal{P}_{\alpha}\big[\partial^{\#}X\big]\,e^{ik_{\rm L}^{(\alpha)}\cdot X(z)}\,\, \bar{\mathcal{P}}_{\alpha}\big[\bar{\partial}^{\#}X\big]\,e^{ik_{\rm R}^{(\alpha)}\cdot X(\bar{z})},$$ 
with $\mathcal{P}_{\alpha}$, $\bar{\mathcal{P}}_{\alpha}$ (to be determined) polynomials and $k_{\rm L}^{(\alpha)}$, $k_{\rm R}^{(\alpha)}$ left- and right-moving  momenta associated to the momentum eigenstate $\alpha$. We wish to derive the explicit form of $V(z,\bar{z})$ and to do so we search for vertex operators which (a) transform correctly under all symmetries of string theory; (b) ideally possess spacetime covariance; (c) are macroscopic and massive; (d) possess classical expectation values, e.g. $\langle X^{\mu}\rangle=X^{\mu}_{\rm cl}$, $\langle J^{\mu\nu}\rangle=J^{\mu\nu}_{\rm cl}$, provided these are compatible with (a), and (e) have small uncertainty in momentum and position (relative to the centre of mass).
Requirement (a) is dictated by string theory, while (b) is preferred for compatibility with standard string technology (e.g. \cite{Polchinski_v1,DHokerPhong}) for string amplitude computations.
Requirements (c-e) would be our targets for a quantum state most closely approximating a large classical string. 

Let us elaborate on (d). Recall that $L_0^{\perp}-\bar{L}_0^{\perp}$ generates rigid spacelike worldsheet translations \cite{Zwiebach04} so that, $\bra{V} [L_0^{\perp}-\bar{L}_0^{\perp},X^i ]\ket{V} = \bra{V} \partial_{\sigma}X^i \ket{V},$ with $L_0^{\perp}$, $\bar{L}_0^{\perp}$ the transverse Virasoro generators (defined below). 
As pointed out in \cite{Blanco-PilladoIglesiasSiegel07}, we see that states invariant under shifts, $(L_0^{\perp}-\bar{L}_0^{\perp})|V\rangle=0$, satisfy $\partial_{\sigma}\langle X^i\rangle=0$, implying that $\langle X^{\mu}\rangle=X^{\mu}_{\rm cl}$ in (d) cannot be realized. This is nevertheless a good condition for classicality when $(L_0^{\perp}-\bar{L}_0^{\perp})|V\rangle\neq0$ and  $\langle X^i\rangle$ is evaluated in lightcone gauge and we will see that this is only possible when the underlying spacetime manifold is compactified in a lightlike direction, $X^-\sim X^-+2\pi R^-$.

For lightcone or covariant gauge states that do not live in a null-compactified background (which satisfy $(L_0^{\perp}-\bar{L}_0^{\perp})|V\rangle=0$ or $(L_0-\bar{L}_0)|V\rangle=0$ respectively), the fact that $\langle X^{\mu}\rangle\neq X^{\mu}_{\rm cl}$ is a gauge problem \footnote{DPS would like to thank Ashoke Sen for very helpful discussions on this.} and says nothing about the classicality of the corresponding quantum states. 
For such states a solution is to fix the gauge completely before evaluating $\langle X^i\rangle$ as was done in \cite{Blanco-PilladoIglesiasSiegel07} but this is somewhat messy and not practical for general states.  Instead, working in lightcone or covariant gauge we shall replace the classicality condition $\langle X^{\mu}\rangle=X^{\mu}_{\rm cl}$ in (d) with \footnote{The authors are very grateful to Joe Polchinski for suggesting this definition of classicality.},
\begin{equation}\label{eq:class_defn}
\big\langle X^i(\sigma',\tau)X^j(\sigma,\tau)\big\rangle =\int_0^{2\pi} \!\!\!\dslash s\,X^i_{\rm cl}(\sigma'-s,\tau)X^j_{\rm cl}(\sigma-s,\tau),
\end{equation}
modulo zero mode contributions, with $X_{\rm cl}^i$ defined in (\ref{eq:Xcl}), $X^i$ given by a similar expression with operators $\alpha_n^i$, $\tilde{\alpha}_n^i,\hat{x}^i,\hat{p}^i$ replacing $\xi_n^i$, $\bar{\xi}^i_n, x^i,k^i$ and $\dslash s \equiv ds/(2\pi)$. Rather than fixing the invariance under $\sigma$-translations on the quantum side (as done in \cite{Blanco-PilladoIglesiasSiegel07}) we average over $\sigma$-translations on the classical side.

We first construct states which satisfy the requirements (a-e). If we proceed by analogy to the harmonic oscillator coherent states, $e^{\lambda a^{\dagger}}|0\rangle$, (with $a|0\rangle=0$ and $[a,a^{\dagger}]=1$) which have classical expectation values, $\partial_t^2\langle x(t)\rangle=-\omega^2 \langle x(t)\rangle$, and consider the naive closed string state $V\sim e^{\lambda_n\cdot \alpha_{-n}}e^{\bar{\lambda}_n\cdot \tilde{\alpha}_{-n}}e^{ik\cdot X(z,\bar{z})}$ we find that the Virasoro constraints are not satisfied \cite{Calucci87}. One possibility is to work in lightcone gauge where the Virasoro constraints are automatically satisfied. Rather than drop spacetime covariance our approach will be to make use of the spectrum generating DDF operators \cite{DelGiudiceDiVecchiaFubini72,AdemolloDelGuidiceDiVecchiaFubini74} which can be used to generate covariant \cite{D'HokerGiddings87,GSW1} physical states.

The DDF operators, $A_n^i$, $\bar{A}_n^i$, satisfy an oscillator algebra, $[A_{n}^i,A_m^{j}]=n\delta^{ij}\delta_{n+m,0}$, in direct analogy to $[\alpha_{n}^{i},\alpha_m^{j}]=n\delta^{ij}\delta_{n+m,0}$. Explicitly, 
\begin{equation}
A^i_n = \oint \dslash z \,\partial X^i\,e^{inq\cdot X(z)},\qquad \bar{A}^i_n = \oint \dslash \bar{z} \,\bar{\partial} X^i\,e^{inq\cdot X(\bar{z})}
\end{equation}
Indices $i$ are transverse to the null vector $q^{\mu}$, $q^2\equiv 0$.
Vertex operators, $V(z,\bar{z})$, have the correct symmetries provided \cite{SasakiYamanaka85} they live in the cohomology $\ker Q/{\rm Im}\, Q$ for all $Q$ in the set of operators $\big\{ L_{n>0},\bar{L}_{n>0},(L_0-1),(\bar{L}_0-1)\big\}$. The DDF operators are gauge invariant, $[L_n,A^i_m]=0$, and so given a physical vacuum, $e^{ip\cdot X(z,\bar{z})}$, for which $Q\cdot e^{ip\cdot X(z,\bar{z})}\cong 0\cong A_{n>0}^i\cdot e^{ip\cdot X(z,\bar{z})}$,  vertex operators of the form $\xi_{i\dots }\bar{\xi}_{j\dots }A_{-n}^{i}\dots \bar{A}_{-\bar{n}}^{j}\dots e^{ip\cdot X(z,\bar{z})}$, are physical and covariant provided $\xi_{\dots i\dots} q^{i}=\bar{\xi}_{\dots i\dots}q^i=0$ and  
\begin{equation}\label{eq:pqcontrs}
p\cdot q=1,\qquad p^2=2\qquad{\rm and} \qquad q^2=0.
\end{equation}
Such vertex operators are transverse to null states (see e.g. \cite{GSW1}) and represent a complete set \cite{D'HokerGiddings87} of covariant vertex operators.

The equivalent lightcone gauge states are obtained by \cite{D'HokerGiddings87} the mapping $A_{-n}^i\rightarrow \alpha_{-n}^i$ and 
$e^{ip\cdot X(z,\bar{z})}\rightarrow |p^+,p^i\rangle$, with $ |p^+,p^i\rangle$ an eigenstate of $\hat{p}^+,\hat{p}^i$ and annihilated by the lowering operators $\alpha_{n>0}^i$, $\tilde\alpha_{n>0}^i$. Here the constraints $(\partial X)^2=(\bar{\partial}X)^2=0$ imply the operator \mbox{equations}
\ben\label{eq:a0abra0}
\al_0^- = \frac{1}{p^+}\left( {L}_0^\perp - 1 \right), \quad
\tilde\al_0^- = \frac{1}{p^+}\left( \bar{L}_0^\perp - 1 \right),
\een
with $L_0^{\perp}$, $\bar{L}_{0}^{\perp}$ the transverse Virasoro generators, ${L}_0^\perp = \frac12 \hat p^i \hat p^i +  N^{\perp}$,  $\bar{L}_0^\perp =\frac12 \hat p^i \hat p^i +  \bar{N}^{\perp}$,
and $N^{\perp}=\sum_{n > 0} {\alpha_{-n}^i\alpha_n^i}$, $\bar{N}^{\perp}=\sum_{n > 0} {\tilde{\alpha}_{-n}^i\tilde{\alpha}_n^i}$.  Recall that $(L_0^{\perp}-\bar{L}_0^{\perp})$ generates spacelike worldsheet shifts. From (\ref{eq:a0abra0}) it follows that, $(\alpha_0^--\tilde{\alpha}_0^-)|V\rangle=\frac{1}{p^+}(L_0^{\perp}-\bar{L}_0^{\perp})|V\rangle$, and so as $\alpha_0^-$ and $\tilde{\alpha}_0^-$ are the left- and right-moving momentum operators, $\hat{p}^-_{\rm L}$ and $\hat{p}_{\rm R}^-$, respectively the lightcone gauge state is only invariant under shifts, $(L_0^{\perp}-\bar{L}_0^{\perp})|V\rangle=0$, when the corresponding eigenvalues, $k_{\rm L,R}^-$, are equal. 


The map between the DDF operators and the lightcone oscillators
suggests that we can define a gauge invariant ``position operator" \cite{GebertNicolai97},
\begin{equation}\label{eq:GIPO}
\mathsf{X}^i(z,\bar{z})-\hat{\mathsf{x}}^i=-i\hat{p}^i\ln |z|^2+i\sum_{n\neq0}\frac{1}{n}\,\big(A_n^i\,z^{-n}+\bar{A}_n^i\,\bar{z}^{-n}\big),
\end{equation}
with $\hat{p}^i=A_0^i=\alpha_0^i$,  $\hat{\mathsf{x}}^i=q_{\mu}J^{i\mu} $ and the angular momentum $J^{\mu\nu} = \oint \dslash z X^{[\mu}\partial X^{\nu]}-\oint \dslash \bar{z} X^{[\mu}\bar{\partial} X^{\nu]}$, integrals being along a spacelike curve, $|z|^2=1$,  and $a^{[\mu\nu]}=\frac{1}{2}(a^{\mu\nu}-a^{\mu\nu})$. Writing $\mathsf{X}^i(z,\bar{z})=\mathsf{X}^i(z)+\mathsf{X}^i(\bar{z})$ and $\hat{\mathsf{x}}^i=\hat{\mathsf{x}}_{\rm L}^i+\hat{\mathsf{x}}^i_{\rm R}$, this satisfies $[L_n,\mathsf{X}^i(z)]=0$ for all $n$, $\big[\mathsf{X}^i(z),\partial_{\tau} \mathsf{X}^j(z')\big]=\delta^{ij}\delta(\sigma-\sigma')$ and similarly for the antiholomorphic piece. Furthermore, $[\hat{\mathsf{x}}^i,\hat{p}^j]=i\delta^{ij}$. Eq.~(\ref{eq:GIPO}) is not essential for what follows but is useful because functionals, $F$, satisfy,
\begin{equation}\label{eq:<cov>=<lc>}
\big\langle F[\mathsf{X}^i(z,\bar{z})-\hat{\mathsf{x}}^i]\big\rangle_{\rm cov}=\big\langle F[X^i(z,\bar{z})-x^i]\big\rangle_{\rm lc},
\end{equation}
which follows from the isomorphism of lightcone (in terms of the $\alpha_n^i,\tilde{\alpha}_n^i$) and covariant states (in terms of the $A_n^i,\bar{A}_n^i$), the isomorphism of the lightcone gauge and gauge invariant position operators, the isomorphism of the corresponding oscillator algebras and finally the fact that the lightcone and covariant states are equivalent. 

Now, a candidate vertex operator to describe bosonic cosmic string loops is the following, 
\begin{equation}\label{eq:VAA}
\begin{aligned}
&V(\lambda,\bar{\lambda})=\\
&C\exp\Big\{\sum_{n=1}^{\infty}\frac{1}{n}\lambda_n\cdot A_{-n}\Big\}\exp\Big\{\sum_{m=1}^{\infty}\frac{1}{m}\bar{\lambda}_m\cdot \bar{A}_{-m}\Big\}\,e^{ip\cdot X(z,\bar{z})}
\end{aligned}
\end{equation}
with $(\lambda,\bar{\lambda})=\{\lambda^i_n,\bar{\lambda}^i_n\}$ and $C=e^{-\sum_{n=1}^{\infty}\left(\frac{1}{2n}|\lambda_n|^2+\frac{1}{2n}|\bar{\lambda}_n|^2\right)}$ a normalization constant. The polarization tensors $\lambda_n^i$, $\bar{\lambda}^i_n$ are such that $\bar{\lambda}_n\cdot q=\lambda_n\cdot q=0$. 

The string theory requirements (see (a-b) above) are  satisfied because any combination of DDF operators on the vacuum yields covariant vertex operators which satisfy the Virasoro constraints. The  cosmic string requirements (c-e) above are also  satisfied: $V(\lambda,\bar{\lambda})$ is an eigenstate of the annihilation operator, $A_{n>0}^i\cdot V\cong \lambda_n^i V,$ and 
hence both $\langle \mathsf{X}^i(\sigma,\tau)-\hat{\mathsf{x}}^i\rangle_{\rm cov}$ 
and  $\langle X^i(\sigma,\tau)-x^i\rangle_{\rm lc}$ 
on account of (\ref{eq:<cov>=<lc>}) are identical to (\ref{eq:Xcl}) with $\lambda_n^i$, $\bar{\lambda}_n^i$ replacing $\xi^i_n$, $\bar{\xi}^i_n$. From the standard coherent state properties it follows that choosing the $|\lambda_n|,|\bar{\lambda}_n|$ appropriately (large) ensures that the cosmic string requirements are satisfied. 

The normal ordered version of (\ref{eq:VAA}) assumes a simple form when $\lambda_n\cdot\lambda_m=\bar{\lambda}_n\cdot\bar{\lambda}_m=0$ (as appropriate for the Burden solutions \cite{Burden85}); in a frame where $\lambda_n\cdot p=\bar{\lambda}_n\cdot p=0$,
\begin{equation}\label{Vnorm}
\begin{aligned}
V(\lambda,\bar{\lambda})&=C\exp\Big\{\sum_{n=1}^{\infty}\frac{1}{n}\lambda_n\cdot P_n(z)e^{-inq\cdot X(z)}\Big\}\\
&\times \exp\Big\{\sum_{m=1}^{\infty}\frac{1}{m}\bar{\lambda}_m\cdot\bar{P}_m(\bar{z})e^{-imq\cdot X(\bar{z})}\Big\}e^{ip\cdot X(z,\bar{z})},
\end{aligned}
\end{equation}
with $P^i_n(z),\bar{P}^i_n(\bar{z})$ related to elementary Schur polynomials, see (\ref {eq:PPbardfn}). This expression follows from bringing the DDF operators close to the vacuum and carrying out the corresponding contour integrals \cite{SklirosHindmarsh09}.

A series expansion of the exponentials shows that we are in fact  superimposing momentum eigenstates with (in general) asymmetric left-right momenta,
$k^{\mu}_{\rm L}-k_{\rm R}^{\mu}=wq^{\mu}$, with winding number $w=N-\bar{N}$ and $q^2=0$. 
Non-zero $w$ and null $q^{\mu}$ implies that the underlying spacetime manifold is null-compactified. 
Any choice of $q^{\mu}$ is permitted provided (\ref{eq:pqcontrs}) and $\bar{\lambda}_n\cdot q=\lambda_n\cdot q=0$ are satisfied. We choose $q^+=q^i= 0$ and $q^-=-R^-$ 
which implies the identification (with $X^+$ non-compact):
\begin{equation*}\label{eq:X^-}
X^-\sim X^-+2\pi R^-.
\end{equation*}
In the rest frame, $k^i=0$, the constraints (\ref{eq:pqcontrs}) lead to: $k^0=\frac{1}{\sqrt{2}}\big(\frac{1}{R^-}+\frac{m^2R^-}{2}\big)$,
 $k^D=\frac{1}{\sqrt{2}}\big(\frac{1}{R^-}-\frac{m^2R^-}{2}\big),$ with $k^{\mu}=\frac{1}{2}(k_{\rm L}+k_{\rm R})^{\mu}$ and mass squared $m^2=N+\bar{N}-2$. The full vertex, $V(\lambda,\bar{\lambda})$, has an effective mass given by 
$\langle m^2\rangle = \langle N\rangle+\langle\bar{N}\rangle-2,$  with $\langle N\rangle=\sum_{n=1}^{\infty}|\lambda_n|^2$ and $\langle\bar{N}\rangle = \sum_{n=1}^{\infty}|\bar{\lambda}_n|^2$. There are similar expressions to $k^0,k^D$ for $\langle k^0\rangle$, $\langle k^D\rangle$ with $\langle m^2\rangle$ replacing $m^2$.

Lightlike compactification in lightcone gauge shows up as follows. Here the states equivalent to (\ref{eq:VAA}) are composed of momentum eigenstates $\al_{-n}^i\dots \al_{-m}^j\tilde{\al}_{-\bar{n}}^k\dots \tilde{\al}_{-\bar{m}}^l\ket{p^+,p^i}$, with $N\neq \bar{N}$ generically (note that $p^i=p^i_{\rm L}=p^i_{\rm R}$). Therefore, from (\ref{eq:a0abra0}) it follows that these states are not translation invariant and $k_L^- \neq k_{\rm R}^-$, thus implying a compact $X^-$ direction. The covariant vertex (\ref{Vnorm}) however is still invariant under shifts and so even though $\langle X^i\rangle_{\rm lc}=X^i_{\rm cl}$ we have $\langle X^i\rangle_{\rm cov}\neq X^i_{\rm cl}$.

Although the above states (\ref{Vnorm}) satisfy the requirements (a-e), the necessity of a null compactified spacetime manifold is perhaps too constraining and so we next discuss the construction of cosmic strings in non-compact spacetimes. Define a projection operator, 
$$
G_w= \int_0^{2\pi}\!\!\!\dslash s\,e^{ is(\hat{w}-w)},
\quad{\rm with}\quad \hat{w}=\hat{p}_{\rm L}^+\hat{p}_{\rm L}^--\hat{p}_{\rm R}^+\hat{p}_{\rm R}^-,
$$
and $\hat{p}_{\rm L}^{\mu}=\oint \dslash z\partial X^{\mu}$, $\hat{p}_{\rm R}^{\mu}=-\oint \dslash \bar{z}\bar{\partial} X^{\mu}$. $\hat{w}$ is the null winding number operator.
This satisfies $G_nG_m=\delta_{n,m}G_n$ and when applied to arbitrary vertices projects out all states in the underlying Hilbert space except for those with null winding number $w$. When there are no transverse compact directions, $\hat{w}=-p\cdot \big(\hat{p}_{\rm L}-\hat{p}_{\rm R}\big)$, with $p^{\mu}$ defined in (\ref{eq:pqcontrs}). Covariant vertex operators in non-compact spacetimes are therefore given by $V_0(\lambda,\bar{\lambda})\cong G_0\cdot V(\lambda,\bar{\lambda})$. With $V(\lambda,\bar{\lambda})$ as given in (\ref{eq:VAA}) we commute $G_0$ through the DDF operators using the expression $e^{is\hat{w}}e^{\sum_{n=1}^{\infty}\frac{1}{n}\lambda_n\cdot A_{-n}}=e^{\sum_{n=1}^{\infty}\frac{1}{n}e^{ins}\lambda_n\cdot A_{-n}}e^{is\hat{w}}$ with a similar relation for the anti-holomorpic sector. This can be derived from the Baker-Campbell-Hausdorff formula, the commutators, $\big[\hat{w},A_{-n}^i\big]=nA^i_{-n}$, $\big[\hat{w},\bar{A}_{-n}^i\big]=-n\bar{A}^i_{-n}$, and the elementary Schur polynomial representation (\ref{eq:S_m(a)a}) with $a_s=\frac{1}{s!}\sum_{n=1}^{\infty}(ins)^{s}\frac{1}{n}\lambda_n\cdot A_{-n}$. This leads us to suggest that the resulting vertex operators represent arbitrary classical loops in non-compact spacetime,
\begin{equation}\label{V_0A}
\begin{aligned}
V_0&(\lambda,\bar{\lambda})=\mathcal{C}_{\lambda\bar{\lambda}}\int_0^{2\pi}\!\!\!\dslash s\exp\Big\{\sum_{n=1}^{\infty}\frac{1}{n}\zeta_n(s)\cdot A_{-n}\Big\}\\
&\quad\qquad\times\exp\Big\{\sum_{m=1}^{\infty}\frac{1}{m}\bar{\zeta}_m(s)\cdot \bar{A}_{-m}\Big\}\,e^{ip\cdot X(z,\bar{z})}
\end{aligned}
\end{equation}
with $\zeta_n^i(s) \equiv \lambda^i_n\,e^{ins}$, $\bar{\zeta}^i_n(s) \equiv \bar{\lambda}^i_n\,e^{-ins}$ and the normalization constant $\mathcal{C}_{\lambda\bar{\lambda}}=\big[\int_{0}^{2\pi}\!\! \dslash s\exp(\sum_{n=1}^{\infty}\frac{1}{n}|\lambda_{n}|^2e^{ins}+\frac{1}{n}|\bar{\lambda}_{n}|^2e^{-ins})\big]^{-1/2}$. Although we do not do so here one can show that this is a coherent state, the definition of which is given in a footnote with unit operator $\mathds{1}=G_0$. The normal ordered version of $V_0(\lambda,\bar{\lambda})$ analogous to (\ref{Vnorm}) can be derived from (\ref{Vnorm}) by computing the operator product $G_0\cdot V(\lambda,\bar{\lambda})$. One finds an expression identical to (\ref{V_0A}) with $P_n^i(z)e^{-inq\cdot X(z)}$, $\bar{P}_n^i(\bar{z})e^{-inq\cdot X(\bar{z})}$ replacing $A_{-n}^i$, $\bar{A}_{-n}^i$ respectively,
\begin{equation*}\label{V_0P}
\begin{aligned}
V_0&(\lambda,\bar{\lambda})=\mathcal{C}_{\lambda\bar{\lambda}}\int_0^{2\pi}\!\!\!\dslash s\exp\Big\{\sum_{n=1}^{\infty}\frac{1}{n}\zeta_n(s)\cdot P_n^i(z)e^{-inq\cdot X(z)}\Big\}\\
&\quad\times\exp\Big\{\sum_{m=1}^{\infty}\frac{1}{m}\bar{\zeta}_m(s)\cdot \bar{P}_m^i(\bar{z})e^{-imq\cdot X(\bar{z})}\Big\}\,e^{ip\cdot X(z,\bar{z})}.
\end{aligned}
\end{equation*}

Having projected out the null winding states worldsheet translation invariance is restored and according to the above discussion the condition for classicality $\langle X\rangle=X_{\rm cl}$ in (d) is replaced by (\ref{eq:class_defn}). Given that we know the classical solution in lightcone gauge, see (\ref{eq:Xcl}), we establish (\ref{eq:class_defn}) for the projected states in lightcone gauge by making use of (\ref{eq:<cov>=<lc>}). Denoting states with null winding $w$ by $V_w(\lambda,\bar{\lambda})\cong G_w\cdot V(\lambda,\bar{\lambda})$ one can show that (\ref{eq:class_defn}) is satisfied by making use of  equations (\ref{eq:Xcl}) and (\ref{eq:GIPO}), with $A_n^i |V_0\rangle=\lambda_n^i|V_n\rangle$, $\bar{A}_n^i |V_0\rangle=\bar{\lambda}_n^i|V_n\rangle$ ($n>0$) and $\langle V_n|V_m\rangle = \delta_{n,m}$, which follow from the DDF operator commutation relations. We learn that (\ref{V_0A}) has a classical interpretation given by (\ref{eq:Xcl}) with $(\xi,\bar{\xi})=(\lambda,\bar{\lambda})$ and $k^i=p^i$.

Finally we show that the angular momentum, $J^{ij}$ defined below (\ref{eq:GIPO}), of the states (\ref{V_0A}) matches the corresponding classical expression. This is a gauge invariant operator and so one expects to find that,
\begin{equation}\label{eq:Jcorr}
\langle J^{ij}\rangle_{\rm cov}=\langle J^{ij}\rangle_{\rm lc}= J_{\rm cl}^{ij},
\end{equation}
We focus on the non-zero mode contribution. The classical expression $J_{\rm cl}^{ij}$ is evaluated using (\ref{eq:Xcl}). The quantity $\langle J^{ij}\rangle_{\rm lc}$ is evaluated using (\ref{eq:<cov>=<lc>}) and the properties used above to establish (\ref{eq:class_defn}). Defining \cite{CornalbaCostaPenedonesVieira06} $B^{n}_{m} =-i \oint \dslash z\,z^{m-1}\,e^{inq\cdot X(z)}$, the quantity $\langle J^{ij}\rangle_{\rm cov}$ is evaluated using the commutation relations $\big[\alpha_m^{i},A^j_n\big] = m\delta^{i,j}B^n_{\phantom{a}m}$, $[A^i_n,B^m_{\ell}\big]=0$, the operator product $B^{-n}_{m}\cdot e^{ip\cdot X(0)}\cong S_{n-m}(nq;0)\,e^{i(p-nq)\cdot X(0)}$, and the property $(B^n_m)^{\dagger}=B^{-n}_{-m}$. All three computations lead to the same result, $\sum_{n>0}\frac{2}{n}{\rm Im}\big(\lambda_n^{*i}\lambda_n^j+\bar{\lambda}_n^{*i}\bar{\lambda}_n^j\big)$, thus establishing statement (\ref{eq:Jcorr}). This expression holds true for the null winding states (\ref{eq:VAA}) as well.

To conclude, we have constructed closed string coherent state vertex operators (in covariant and lightcone gauge) and have shown how to map these to arbitrary classical solutions. Given that we know the classical solutions in lightcone gauge we make use of (\ref{eq:<cov>=<lc>}) to extract the lightcone gauge position expectation values from the equivalent covariant gauge states. We found that the naive covariant (or  lightcone) gauge construction  (\ref{eq:VAA}) is only consistent when the underlying spacetime manifold is compactified in a lightlike direction. Here we note that: (1) the string only fluctuates in directions transverse to the null direction implying that the various geometrical features of the string (such as cusps) are not affected by the compactification, (2) the mass spectrum for states in the null-compactified Hilbert space is as in the non-compact case, $m^2=N+\bar{N}-2$, but with $N$ not necessarily equal to $\bar{N}$, (3) expectation values $\langle X^{\mu}\rangle$ for $\mu=(\pm,i)$ are also as in the non-compact case when $\langle N\rangle =\langle \bar{N}\rangle$, implying that classically compact and non-compact $X^-$ are indistinguishable. Quantizing on a null compact background is known as Discrete Lightcone Quantisation (DLCQ) \cite{PauliBrodsky85}, and is a crucial component in the M(atrix) theory to string theory correspondence \cite{Susskind97}. 

We then discussed the construction of classical cosmic string loops in non-compact spacetimes and showed that these satisfy the requirements (a-e) when the definition for classicality $\langle X\rangle=X_{\rm cl}$ in (d) is replaced by (\ref{eq:class_defn}). Finally, we showed that the angular momenta of the covariant, lightcone gauge and classical descriptions are identical (\ref{eq:Jcorr}) thus providing further support for the conjecture that arbitrary classical string solutions (\ref{eq:Xcl}) are described in string theory by the covariant (or corresponding lightcone gauge) states (\ref{V_0A}).

These new vertex operators may be used to study the cosmic string evolution predicted by string theory, taking gravitational backreaction into account which is almost always neglected in the classical computations. It can also be used to check whether gravitational radiation is indeed the primary decay channel of cosmic strings, and if so what the frequency spectrum is. 

\textit{Appendix:} Elementary Schur polynomials are defined  \cite{Kac90} by the generating series, $\sum_{m=0}^{\infty}S_m(a_1,\dots,a_m)z^m=\exp \sum_{n=1}^{\infty}a_n\,z^n
$ and read explicitly:
\begin{subequations}\label{eq:S_m(a)}
\begin{align}
S_m(a_1,\dots,a_m)&=\sum_{k_1+2k_2+\dots+mk_m=m}\frac{a_1^{k_1}}{k_1!}\dots\frac{a_m^{k_m}}{k_m!}\label{eq:S_m(a)a}\\
&=-i\oint_0\dslash w\,w^{-m-1} \exp \sum_{s=1}^{m}a_sw^s\label{eq:S_m(a)b}
\end{align}
\end{subequations}
with $\dslash w\equiv dw/(2\pi)$, $S_0=1$ and $S_{m<0}=0$. When $a_s=-\tfrac{1}{s!}inq\cdot\partial^sX(z)$, with $q^{\mu}$ defined in  (\ref{eq:pqcontrs}) we write $S_m(nq;z)\equiv S_m(a_1,\dots,a_m)$. The following Taylor series is useful, $e^{-inq\cdot X(z)}=\sum_{a=0}^{\infty}z^aS_a{(nq;0)}e^{-inq\cdot X(0)}$. The polynomials $P_n(z)$, $\bar{P}_n(\bar{z})$ that appear in the normal ordered covariant coherent state (\ref{Vnorm}) are then defined by 
\begin{subequations}\label{eq:PPbardfn}
\begin{align}
P_n^i(z) &= \sum_{\ell=1}^{\infty}\frac{i}{(\ell-1)!}\,\partial^{\ell}X^i(z)S_{n-\ell}(nq;z),\label{eq:Pn}\\
\bar{P}_n^i(\bar{z}) &= \sum_{\ell=1}^{\infty}\frac{i}{(\ell-1)!}\,\bar{\partial}^{\ell}X^i(\bar{z})\bar{S}_{n-\ell}(nq;\bar{z}).\label{eq:Pnbar}
\end{align}
\end{subequations}

\textit{Acknowledgements}: The authors are gratefully indebted to Joe Polchinski for providing crucial insight and suggestions which ultimately made the coherent state construction in non-compact spacetime possible. Both authors have benefited greatly from discussions with Diego Chialva. DS would also like to thank Jose Blanco-Pillado, Ed Copeland, Andy Strominger, Tanmay Vachaspati and especially Ashoke Sen for very helpful discussions and suggestions.

\bibliography{spi}

\end{document}